\documentclass[12pt]{article}
\usepackage[dvips]{graphicx}
\usepackage{amsmath}

\begin {document}
\begin{flushright}
{\small
SLAC--PUB--11375\\
July 2005}
\end{flushright}

\vspace{20pt}

{
\begin{center}
{{\bf\LARGE Nearly Conformal QCD and AdS/CFT}\footnote{Work supported by the Department
of Energy contract DE--AC02--76SF00515.}}\\

\vspace{20pt}

{\bf Guy F. de T\'eramond\footnote{E-mail: gdt@asterix.crnet.cr}
and Stanley J. Brodsky\footnote{E-mail: sjbth@slac.stanford.edu} } \\

\vspace{10pt}

{\it Universidad de Costa Rica, San Jos\'e, Costa Rica\\ \vspace{10pt}
Stanford Linear Accelerator Center, \\ 
Stanford University, Stanford, California 94309, USA 
 }
\end{center}
}

\vspace{40pt}

\begin{abstract}

The AdS/CFT correspondence is a powerful tool to study the properties of conformal
QCD at strong coupling in terms of a higher dimensional dual gravity theory. 
The power-law falloff of scattering amplitudes in the non-perturbative regime and calculable
hadron spectra follow from holographic models dual to QCD with conformal behavior at short
distances and confinement at large distances. String modes and fluctuations 
about the AdS background are identified with QCD degrees of freedom and orbital
excitations at the AdS boundary limit. A description of form factors in space and time-like
regions and the behavior of light-front wave functions can also be understood in terms of a dual 
gravity description in the interior of AdS.

\end{abstract}

\vspace{20pt}

\begin{center}
{\it Presented at  \\
First Workshop on Quark-Hadron Duality and the Transition to pQCD\\
6-8 June 2005  \\
Frascati, Italy \\
 }
\end{center}

\vfill

\newpage

\parindent=1.5pc
\baselineskip=16pt

\setcounter{footnote}{0}

The correspondence~\cite{Maldacena:1997re}, between string theory in a warped 10-dimensional
space and conformal Yang-Mills theories defined at its four dimensional space-time
boundary at infinity has led to important insights into the properties
of QCD at strong coupling. The applications include the nonperturbative 
derivation~\cite{Polchinski:2001tt}
of dimensional counting rules~\cite{Brodsky:1973kr}
for hard exclusive glueball scattering, the description
of deep inelastic structure functions~\cite{Polchinski:2002jw} and the
power falloff of hadronic light-front wave functions (LFWF) including orbital angular
momentum~\cite{Brodsky:2003px}.
In the original correspondence~\cite{Maldacena:1997re}, 
the low energy supergravity approximation to type IIB string compactified
on AdS$_5 \times S^5$, is dual to the $\mathcal{N} = 4$ super Yang-Mills (SYM) 
theory at large $N_C$.
 
QCD is fundamentally
different from SYM theories where all of the matter fields transform in adjoint multiplets
of $SU(N_C)$. Its string dual is unknown. We assume however that
such a string exists and that, in principle, it can be defined in terms of the QCD degrees
of freedom at infinity. In practice, we can deduce some of the dual string properties by 
studying its boundary ultraviolet limit, $r \to \infty$, as well as 
the small-$r$ infrared region of AdS space, characteristic of strings 
dual to confining gauge theories. This approach,
which can be described as a bottom-up approach, has been successful in obtaining general properties of
the low-lying hadron spectra, chiral symmetry breaking and hadron couplings in 
AdS/QCD~\cite{Boschi-Filho:2002ta}, in addition
to the scattering results described in~\cite{Polchinski:2001tt,Polchinski:2002jw,Brodsky:2003px}.
A different approach, a top-bottom approach, consists in studying the 
full supergravity equations to compute the glueball
spectrum~\cite{Csaki:1998qr}. The addition of D3/D7 branes~\cite{Karch:2002xe} or the
gravitational fluctuations of a thick brane in Minkowski space~\cite{Siopsis:2005yj}
leads to a calculable meson spectrum. 
Other aspects of high energy scattering
in warped spaces have been addressed in~\cite{Giddings:2002cd}.

It is remarkable that dimensional scaling for exclusive processes works so well at 
relatively low energies where higher twist effects are expected to be 
dominant~\cite{FrascatiProc:2005}. Counting rules can be understood if 
QCD resembles a strongly
coupled theory at moderate energies. The isomorphism of the group $SO(2,4)$, which act as the group of
conformal symmetries at the AdS boundary in the limit of massless quarks and vanishing 
$\beta$ function, 
with the group of isometries of AdS,
$x^\mu \to \lambda x^\mu$, $r \to r / \lambda$,
maps  scale transformations into the holographic coordinate $r$.
Consequently, the string mode in $r$ is the extension of the hadron wave function into the fifth
dimension. In particular, the $r \to 0$ boundary corresponds to the zero separation limit between 
quarks. Conversely, color confinement implies that there is a maximum separation of quarks and 
a minimum value $r_0 = \Lambda_{\rm QCD} R^2$, where the string modes can propagate. The cutoff
at $r_0$ is dual to the introduction of a mass gap $\Lambda_{\rm QCD}$, it breaks the conformal 
invariance 
and is responsible for the generation of a spectrum of color singlet hadronic states.

The duality between a gravity theory on 
$AdS_{d+1}$ space and the strong coupling limit of a conformal gauge theory at its $d$-dimensional 
boundary,
is given in terms of the $d+1$ partition function in the bulk
$Z_{grav}[\Phi(x,z)] = \int \mathcal{D}[\Phi] e^{ i S_{grav}[\Phi]}$,
and the $d$-dimensional functional integral over quarks $q$ and gluons $G$ in
presence of an external source,
\begin{equation}
  Z_{QCD}[\Phi_o(x)] = \int [\mathcal{D} G] [\mathcal{D} q]
  \exp{\left\{i S_{QCD}[G,q] + i \int d^dx \Phi_o \mathcal{O} \right\} },
 \label{eq:Zqcd}
 \end{equation}  
with  conditions~\cite{Gubser:1998bc}
 $Z_{grav}\left[\Phi(x,z)_{z = 0} = \Phi_o(x) \right]
 = Z_{QCD}\left[\Phi_o\right]$, where $z = R^2/r$ and $R$ is the AdS radius.  
Near the $AdS$ boundary, $z \to 0$, $\Phi(x,z)$ behaves as
$ \Phi(x,z) \to  z^\Delta \Phi_+(x) + z^{{d - \Delta}} \Phi_-(x)$, 
where  $\Phi_-(x)$ is the boundary source, $\Phi_- = \Phi_o$,
and $\Phi_+(x)$ is the normalizable solution  with conformal dimension $\Delta$. 
The physical string modes 
$\Phi(x,z) \sim e^{-i P \cdot x} f(r)$,
are plane waves along the Poincar\'e coordinates with four-momentum  $P^\mu$
and hadronic invariant mass states $P_\mu P^\mu = \mathcal{M}^2$. 
For large-$r$, $f(r) \sim r^{-\Delta}$. The dimension $\Delta$
of the string mode, is the same dimension of the interpolating operator {\small$\mathcal{O}$}
which creates a hadron out of the vacuum:
$\langle P \vert \mathcal{O} \vert 0 \rangle  \neq 0$.

QCD degrees of freedom  are defined at the AdS boundary at infinity. 
Quarks and gluons also propagate in the AdS interior. In the
limit where the linearized equations for spin 0, $\frac{1}{2}$, 1 and $\frac{3}{2}$ 
on $AdS_5 \times S^5$ have no interactions, only color singlet states
of dimension 3, 4 and $\frac{9}{2}$ have dual string modes and a physical spectrum. 
Consequently, only
the hadronic states  (dimension-$3$) $J^P=0^-,1^-$ pseudoscalar
and vector mesons, 
the (dimension-$\frac{9}{2}$) $J^P=\frac{1}{2}^+, \frac{3}{2}^+$ 
baryons, and the (dimension-$4$) $J^P= 0^+$ glueball states,  
corresponding exactly to the lowest-mass physical hadronic states can be derived in the
classical holographic limit~\cite{deTeramond:2005su}. Hadrons fluctuate in
particle number, but there are also orbital angular momentum fluctuations. A major difficulty
in describing the hadron spectrum with AdS/CFT arises from the nature of the string solutions,
since duality cannot be established for spin $>$ 2, where the conformal dimensions become 
very large.
Higher Fock components are manifestations of the
quantum fluctuations of QCD; metric
fluctuations of the bulk geometry about the fixed AdS background should
correspond to quantum fluctuations of Fock states above the valence state.
Indeed, for large Lorentz spin,
orbital excitations in the boundary
correspond to quantum fluctuations about the AdS
metric~\cite{Gubser:2002tv}. We identify the higher spin hadrons with the
fluctuations around the spin 0, $\frac{1}{2}$, 1 and $\frac{3}{2}$ classical
string solutions on the $AdS_5$ sector~\cite{deTeramond:2005su}. 

As a specific example, consider the twist (dimension minus spin) two glueball 
interpolating operator
$\mathcal{O}_{4 + L}^{\ell_1 \cdots \ell_m} = F D_{\{\ell_1} \dots D_{\ell_m\}} F$ with
total internal space-time orbital
momentum $L = \sum_{i=1}^m \ell_i$ and conformal dimension $\Delta_L = 4 + L$.
We match the large $r$ asymptotic behavior of each string mode
to the corresponding
conformal dimension of the boundary operators
of each hadronic state while maintaining conformal invariance.
In the conformal limit, an $L$ quantum, which is
identified with a quantum fluctuation about the AdS geometry,
corresponds to an effective five-dimensional mass $\mu$ in
the bulk side.  The allowed values of $\mu$ are uniquely determined by
requiring that asymptotically the dimensions become spaced by
integers, according to the spectral relation 
$(\mu R)^2 = \Delta_L(\Delta_L - 4)$~\cite{deTeramond:2005su}.
The interaction term in Eq. \ref{eq:Zqcd} for a state
with orbital $L$ at the asymptotic boundary results
in the effective coupling
\begin{equation}
 S_{int} = \int d^4 x~ \partial_{x_{\ell_1}} \cdots  \partial_{x_{\ell_m}} 
 \Phi(x,z)\vert_{z=0} ~\mathcal{O}^{\ell_1 \cdots \ell_m} .
\end{equation}
The string modes
$\Phi^{\ell_1 \dots \ell_m}(x,z) =  \partial_{x_{\ell_1}} \cdots  \partial_{x_{\ell_m}}
\Phi(x,z)$ for a given eigenvalue $\mu R$, with $\Phi(x,z) = C e^{-i P \cdot x} 
z^2 J_{\alpha}(z \beta_{\alpha,k} \Lambda_{QCD})$, $\alpha = \Delta_L -2$, have the
correct Lorentz structure at the AdS boundary, since each $\partial_{x^\mu}$  
pulls down a $P_\mu$ from the exponential factor of the string mode, leaving intact the holographic
$z$-dependence which is determined  by the conformal dimension $\Delta_L$. 

\begin{figure}[h]
\centering
\includegraphics[width=10.0cm]{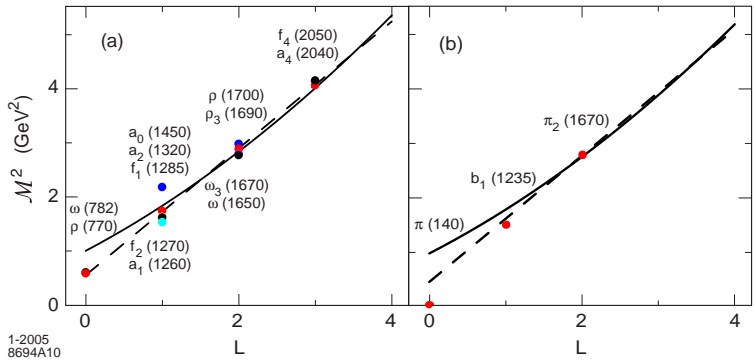}   
\caption{\small Light meson orbital states for $\Lambda_{\rm QCD} = 0.263$ GeV: (a) vector mesons and
(b) pseudoscalar mesons. The dashed line 
is a linear Regge trajectory with slope 1.16 ${\rm GeV}^2$.}
\label{fig:MesonSpec}
\end{figure}

The four-dimensional mass spectrum follows from the Dirichlet boundary condition  $\Phi(x,z_o) = 0$,
$z_0 = 1 / \Lambda_{\rm QCD}$, on 
the AdS string modes for the different wave functions corresponding to spin $<$ 2 and is
given in terms of the zeros of Bessel functions, $\beta_{\alpha,k}$. In the case of 
mesons the predicted spectrum
is shown in Figure \ref{fig:MesonSpec} for $\Lambda_{\rm QCD} = 0.263$ GeV, the only parameter in
the model. The baryon spectrum is discussed in~\cite{deTeramond:2005su}. 

\begin{figure}[h]
\centering
\includegraphics[width=6cm]{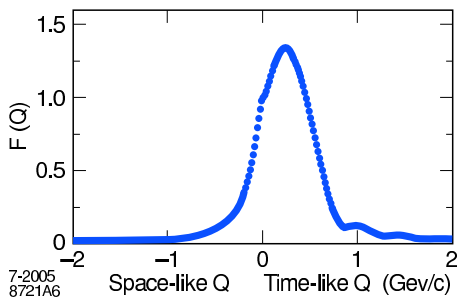}
\caption{\small Space and time-like structure for the pion form factor in AdS/QCD.}
\label{fig:FF}
\end{figure}

The form factor in AdS/QCD is the overlap of the
normalizable modes dual to the incoming
and outgoing hadron $\Phi_I$ and $\Phi_F$ and the
non-normalizable mode $J(Q,z)$, dual to the external source
\begin{equation}
F(Q^2)_{I \to F} 
\simeq R^{3 + 2\sigma} \int_0^{z_o} \frac{dz}{z^{3 + 2\sigma}}~
 \Phi_F(z)~J(Q,z)~\Phi_I(z),
\label{eq:FF}
\end{equation}
where $\sigma_n = \sum_{i=1}^n \sigma_i$ is the spin of the interpolating
operator $\mathcal{O}_n$, which creates an $n$-Fock state $\vert n \rangle$ at the
AdS boundary. $J(Q,z)$  has the value 1 at
zero momentum transfer, and
as boundary limit the external current, thus
$A^\mu(x,z) = \epsilon^\mu e^{i Q \cdot x} J(Q,z)$. The solution
to the AdS wave equation subject to  boundary conditions at  $Q = 0$ and
$z \to 0$ is~\cite{Polchinski:2002jw}  $J(Q,z) = z Q K_1(z Q)$.
At large enough $Q \sim r/R^2$, the important contribution to (\ref{eq:FF}) is
from the region near $z \sim 1/Q$.
At small $z$, the $n$-mode $\Phi^{(n)}$ scales as 
$\Phi^{(n)} \sim z^{\Delta_n}$, and we recover the power law scaling~\cite{Brodsky:1973kr},
$F(Q^2) \to \left[1/Q^2\right]^{\tau_n - 1}$,
where the twist $\tau_n = \Delta_n - \sigma_n$, 
is equal to the number of partons, $\tau_n = n$.
A numerical computation for the pion form factor~\cite{SJBGdT:2005} gives the results 
shown in Figure \ref{fig:FF}, where the resonant structure in the time-like region from the 
AdS cavity modes is apparent.

\begin{figure}[h]
\centering
\includegraphics[width=12.8cm]{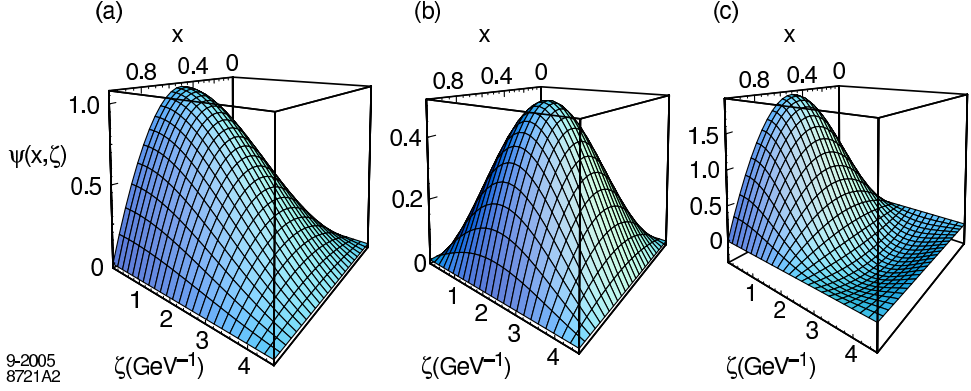}
\label{fig:AdSLFWF}
\caption{\small Two-parton bound state holographic LFWF $\psi(x, \zeta)$: 
(a) ground state $\ell = 0$, $k = 1$, (b) first orbital excited
state $\ell = 1$, $k = 1$, (c) first radial exited state $\ell = 0$, $k = 2$.}
\label{fig:LFWF}
\end{figure}

The AdS/QCD correspondence provides also a simple description of hadrons at the amplitude level
by mapping string modes to the impact space representation of LFWF. 
In terms of the partonic variables
$ x_i \vec r_{\perp i} = \vec R_\perp + \vec b_{\perp i}$,
where $\vec r_{\perp i}$ are the physical transverse position coordinates,  
$\vec b_{\perp i}$ internal coordinates, 
$\sum_i \vec b_{\perp i} = 0$,  and $\vec R_\perp$ the
hadron transverse center of momentum
$\vec R_\perp = \sum_i x_i \vec r_{\perp i}$,
$\sum_i x_i = 1$, we find for a two-parton LFWF 
\begin{equation*}
\psi_{\ell,k}(x,\zeta) = C
 x(1-x) J_{1 + \ell}\left(\zeta \beta_{1+\ell,k} \Lambda_{QCD}\right)/\zeta,
\end{equation*}
where $\zeta = \vert \vec b_\perp \vert \sqrt{x(1-x)}$ represents the scale of the invariant
separation between quarks.
The first eigenmodes are depicted in Figure \ref{fig:LFWF}.

The holographic model is quite successful in describing the known light
hadron spectrum. The only mass  scale is $\Lambda_{QCD}$. The model incorporates confinement
and conformal symmetry. Only dimension  $3, \frac{9}{2}$ and 4 states $\bar q q$, $q q q$,  
and  $g g$ appear in the duality at the classical level. 
Non-zero orbital angular momentum and higher Fock-states require the introduction of quantum 
fluctuations. The model gives a
simple description of the structure of hadronic form factors and LFWFs.
It explains the suppression of the odderon. The
dominance of quark-interchange in hard exclusive processes emerges naturally from the classical
duality of the holographic model.

\end{document}